\documentclass[%
superscriptaddress,
 amsmath,amssymb,
 aps,prfluids,
]{revtex4-2}
\usepackage{color, float}
\usepackage{graphicx}
\usepackage{dcolumn}
\usepackage{bm}

\begin{document}

\preprint{APS/123-QED}

\title{How does gravity influence freezing dynamics of drops on a solid surface}

\author{Hao Zeng}
\author{Sijia Lyu}
\affiliation{Center for Combustion Energy, Key Laboratory for Thermal Science and Power Engineering of Ministry of Education,
Department of Energy and Power Engineering, Tsinghua University, 100084 Beijing, China}
\author{Dominique Legendre}
\affiliation{Institut de Mécanique des Fluides de Toulouse (IMFT), Université de Toulouse, CNRS -Toulouse, France}
\author{Chao Sun }
 \thanks{chaosun@tsinghua.edu.cn}
\affiliation{Center for Combustion Energy, Key Laboratory for Thermal Science and Power Engineering of Ministry of Education,
Department of Energy and Power Engineering, Tsinghua University, 100084 Beijing, China}
\affiliation{Department of Engineering Mechanics, School of Aerospace Engineering, Tsinghua University, Beijing 100084, China
}

\date{\today}

\begin{abstract}

Water droplet freezing is a common phenomenon in our daily life. In both natural scenarios and industrial production, different surface inclinations bring distinctive deformation and freezing dynamics to frozen droplets. We explore the freezing of pendent and sessile droplets at different Bond number regimes. The effect of gravity on the droplet freezing process is analyzed by considering droplet morphology, freezing front dynamics, and freezing time. It is found that gravity can significantly influence droplet freezing processes via shaping the initial droplet, resulting in the flattening or elongation of pendent and sessile droplets, respectively. We show that the droplet initial geometry is the most important parameter and it completely controls the droplet freezing. Despite the significant difference in the initial droplet shape several remarkable similarities have been found for pendent and sessile droplets at small and large Bond numbers.
The final height of a frozen droplet is found to be linearly proportional to its initial height. The time evolution of the ice-liquid-air contact line is found to reproduce the power-law $t^{0.5}$, but noticeably faster than the Stefan 1-D icing front propagation. As a consequence, the time to freeze a droplet is faster than predicted by the Stefan model and it is found to be dependent on the initial droplet height and base radius through a simple power-law. 
\end{abstract}

\maketitle


\section{\label{sec:level1}Introduction}

Water droplet freezing is an important and commonly seen phenomenon in our daily life, especially in cold-weather regions. This process also significantly affects many important industrial applications. In the power delivery industry, the accumulated ice could cause damage to the electrical wires and transformers\cite{szilder2002study}. In the aerospace industry, ice formation on the aircraft wings is a central issue that may increase fuel consumption and threaten aviation safety\cite{lynch2001effects}. In the wind power plant, the ice accretion on the wind turbine blade will drastically increase loads and reduce the system performance\cite{makkonen2001modelling}. On the other hand, the solidification of liquid droplets is also dominant in many additive manufacturing processes, such as the metal alloy droplet deposition\cite{adaikalanathan2019numerical}, micro-droplet 3D printing\cite{chen2020research}, and laser welding\cite{Reddy_2018}. The solidification dynamics and morphology would lead to different microstructures and divergence of material characteristics. To further understand the mechanisms governing this very common phenomenon, prevent undesirable ice formation and better utilizing droplet solidification in productions, it is important to study the freezing features of water droplets.
 
The freezing processes are the focus of many current studies in a broad scope. {Since the first reported shapes of freezing droplets half a century ago\cite{Stairs1971}, the community has been trying to find the conditions that would affect the droplet solidification process from many different aspects.} The substrate properties are proved to be one of the most important aspects that would affect the freezing process via substrate temperature\cite{ghabache2016frozen,thievenaz2020retraction} and wettability\cite{zhang2016freezing,liu2017distinct,dong2019quasi,jin2017impact,schremb2017transient}. The environment medium is also taken into consideration of the main factors that influence the freezing process. The still gaseous medium is always considered as an adiabatic condition for its low thermal conductivity\cite{Marin2014}, while the interface evolution would be greatly different when flow exists in the medium\cite{vu2018fully, Jung2012,roisman2015dislodging,lian2017experimental}. The properties related to the droplet itself are even much more complex. The initial conditions of the droplet as well as the ambiant condition such as humidity\cite{Sebilleau2021}  largely determine the subsequent process of the freezing kinetics. The supercooling \cite{zhang2018simulation} and impingment\cite{thievenaz2020freezing,wang2021numerical,thievenaz2020retraction,jin2016impact}  affect the freezing process by changing the initial properties and morphology of the droplet. Experiments\cite{zhang2016effect,zhang2017modelling} indicate that the drop size is a critical factor that influences the heat change efficiency in the freezing process. Different initial configurations exhibit both universality and diversities in the freezing process\cite{ismail2016universality,thievenaz2020retraction}. The initial configuration of the droplet is determined by the balance of surface tension, the intrinsic property of free liquid surfaces, and  gravitational effects on the Earth. The surface tension force would change with the surface curvature and successively adapt the droplet shape during the freezing process to minimize the surface energy\cite{lyu2021hybrid}. The gravity puts many restrictions in the  experiments on the Earth mainly due to reduced droplet size and detachment from resolidified liquid-solid systems\cite{1963On,jeromen2012modelling}. Several models are proposed to discuss the influence of gravity on the solidification of droplets, but in such studies, the shape and thermal problems are decoupled and density variation is not considered\cite{sanz1987influence} for simplicity. The effect of gravity in droplet freezing process has still not been well studied. 

In this study, we mainly investigate the influence of gravity in water drops freezing on a copper plate cooled down below 0 $^\circ C$. After a short description of the experimental setup and numerical model, we qualitatively describe the freezing processes of pendent droplets and sessile droplets. The gravitational effects on droplet morphology and freezing dynamics are extensively explored with experimental and numerical results. Further more, several remarkable similarities are found for pendent and sessile droplets, and for both small and large Bond numbers. The results present intriguing laws in the droplet freezing process with different gravity directions.

\section{\label{sec:2-level1} Method}

\subsection{\label{sec:2-level2} Experiment setup}

\begin{figure*}
\includegraphics[width=0.7\textwidth]{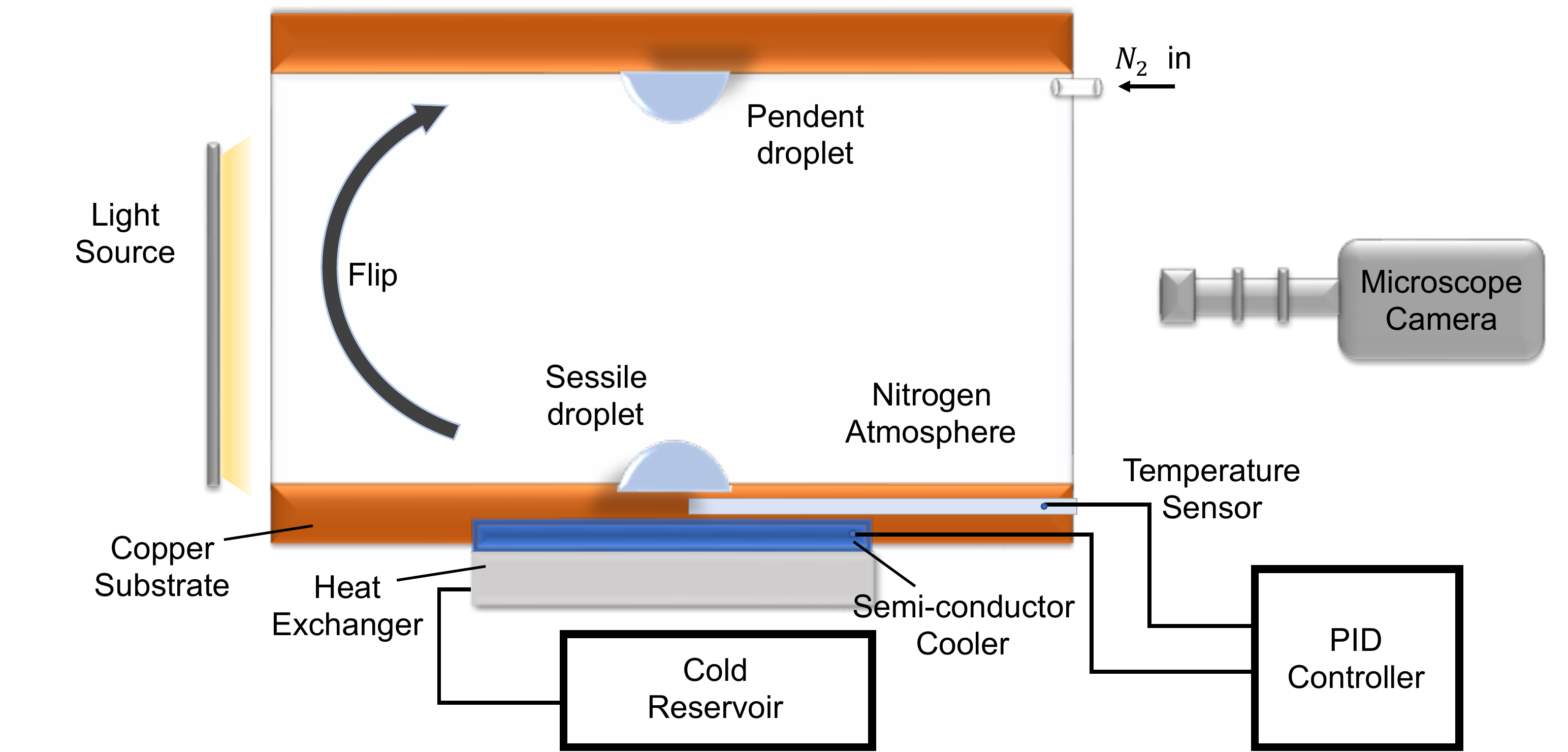} 
\caption{\label{fig:setup} Sketch of the experiment setup. The setup is flexible to turnover, thus pendent droplet and sessile droplet can be studied with the same apparatus }
\end{figure*}
The gravitational effect on the freezing of singular water droplets can be studied with the experimental setup shown in Fig.~\ref{fig:setup}. In the experiment, distilled water droplets are cooled on a copper substrate. A pipette is used to create single droplet with a certain volume in the experiment. To better control the droplet shape, concave outline is manufactured on the substrate to fix the solid-liquid-vapor contact line at droplet base. Since the concave depth is much less than the length scale of the droplet, the influence of the structure on the freezing process can be neglected. 

A semi-conductor cooler is embedded in the substrate to cool the substrate. A heat exchanger connecting to a cold reservoir is attached to the opposite side of the semiconductor to maintain energy balance. The surface temperature of the substrate is measured with a thermocouple embedded 0.5 mm beneath the upper surface. The thermocouple is connected with a PID controller to adjust the power supply to the semiconductor cooler so that we can obtain a constant surface temperature with a variation of less than 0.2 K. {In the experiments, the substrate is cooled down from room temperature to the desired temperature after droplet deposition. The substrate cooling process takes about 3 minutes, giving an average cooling rate of about 0.3 K per second. After the substrate cooling down, there is sufficient time for the droplet to reach a thermal equilibrium state before the freezing initiation. During droplet freezing, the substrate surface temperature is controlled to $-10$ K below the freezing temperature.} The cooling area of the semi-conductor cooler fully covers the contact region of the droplet with the substrate, thus the surface temperature in contact with the droplet can be assumed uniform during the freezing process.

The experiments are conducted in a highly transparent plexiglass chamber with an inner size of 8 cm $\times$ 8 cm $\times$ 8 cm, and a wall thickness of 5 mm. To prevent frost formation on the freezing substrate, the chamber is filled with pure nitrogen gas at a slight overpressure. At the beginning of every experiment, the chamber is flushed with nitrogen gas, while the flow rate is kept to the minimum during observations to reduce any possible disturbance to the freezing process. The temperature outside of the chamber is consistent with the temperature of the workspace at $298$ K. 

The freezing process is recorded using a high-speed imaging system from the side view with back-light illumination. The imaging system utilizing a high-speed camera (Photron Nova S12) combined with a long-distance microscope (Zoom 6000 by Navitar) results in a maximum resolution of 5.7 $\mu$m/pixel. With this imaging system, we can precisely acquire high-resolution images for analyzing the droplet freezing features versus time.\\

\subsection{\label{sec:level2} Numerical model}

Based on previous studies of the freezing process of droplets \cite{zhang2017modelling,sanz1987influence,Anderson1996}, a simplified phenomenological model is developed in this study. The following widely used assumptions are adopted in the model \cite{zhang2018simulation,Lu2022}: Firstly, the gas-liquid interface is assumed to be axi-symmetric, and the gravity direction coincides with the axis of symmetric of the droplet. Secondly, the solidification front remains flat in the freezing process, and heat transfer in the ice layer is one-dimensional. Thirdly, the inner flow inside the unfrozen liquid is negligible. Fourthly, the evaporation and heat transfer at the gas-liquid interface is neglected. Finally, the temperature of ice-water mixture remains constant in the freezing process. The temperature variation of density and thermal diffusivity are not considered, and the change of properties due to the phase change is assumed to happen suddenly at the solidification front.

\begin{figure}
\includegraphics[width=0.90\textwidth]{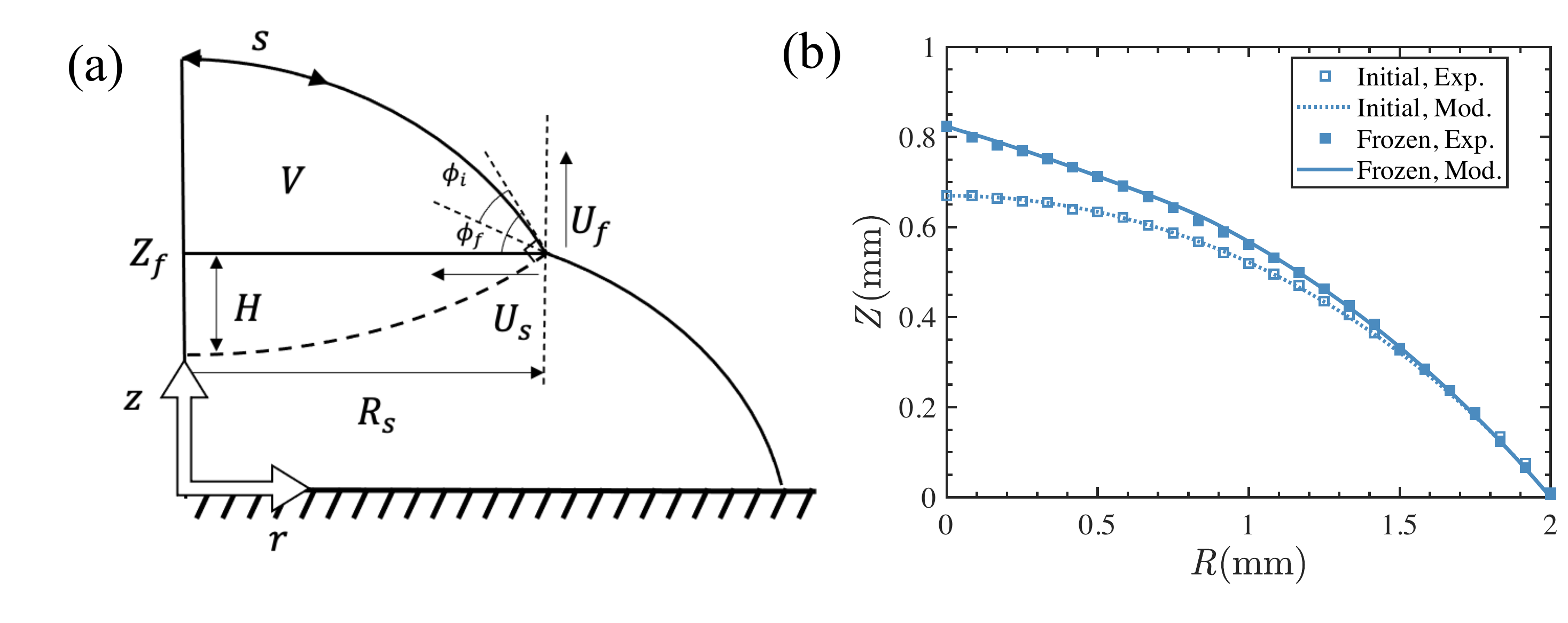} 
\caption{\label{fig:model} (a) A schematic diagram showing the features of the simple model developed in this study to describe the solidification process of a droplet; (b) Experimental droplet profiles compared with the calculation results from the model. The open circle and dashed line represent the initial profile measured in the experiment and calculated from the model, respectively. The solid circle and solid line represent the final profile measured in the experiment and calculated from the model, respectively}
\end{figure}

With these assumptions, the droplet freezing process can be resolved from mass conservation of the liquid-solid system as shown in Fig.~\ref{fig:model}(2). The initial profile as well as surface of the remaining liquid in the freezing process are calculated by integrate the Young-Laplace equation along the liquid-gas interface \cite{Cai2020}. Dilatation is considered at the freezing front by the conservation of total mass and receding contact angle \cite{sanz1987influence}. With the presence of a contact line slip at the late stage of freezing \cite{ANDERSON1994245}, dynamic "growth angle", which is defined as the angle between the tangents to the solid-gas and liquid-gas interface, is implied to the model to predict the movement direction of tri-junction. The thermal problem can be characterized by the energy conservation at the freezing front. Due to the second assumption we take, the temperature is linear distributed between the substrate and the front \cite{Lu2022}, and the front propagation can be described by the solution of one phase Stefan problem at the beginning of the freezing process \cite{Fang2020}. In the later stage of freezing, considering that the freezing front takes spherical shape \cite{Marin2014}, the conducted heat flux is calculated along a virtual spherical cap (as shown in Fig.~\ref{fig:model}(a), the dashed line) confirmed by the freezing front radius and liquid contact angle to the ice. The detailed derivation of the model is provided in the appendix.

\begin{table}[h]
\caption{\label{tab:physical property} Specific values of physical properties used in the calculation.}
\begin{ruledtabular}
\begin{tabular}{lccc}
Parameters & Units & Water & Ice\\\hline
Density & $\rm kg/m^3$ & 1000 & 917\\
Thermal diffusivity& $\rm mm^2/s$ & $0.132$ & $1.176$\\
Heat capacity & $\rm kJ/kg\cdot K$ & / & 2.028\\
Latent heat & $\rm kJ/kg$ & 333.4 & /\\
surface Tension & $\rm N/m$ & 0.0728 & /\\
Gravity & $\rm m/s^2$ & 9.81 & 9.81\\
\end{tabular}
\end{ruledtabular}
\end{table}

With the physical properties given in Table~\ref{tab:physical property} in the appendix, the calculation results are compared to experiments to validate the reliability of this model. Figure~\ref{fig:model}(b) shows typical initial and final profiles obtained experimentally. These  data correspond to a droplet with supercooling, which is the temperature difference between the cold wall and the freezing point, $\Delta T = T_m -T_w = 10 \ \rm K$ and volume of $V_0= 7.8 \ \rm{\mu L}$ in different directions of gravity. Through experiments, we can get appropriate values of $V_0$ to calculate the theoretical droplet profile with the model. Good agreements are found between the theoretical and experimental results for both the initial and final droplet profile.

\color{black}
\section{\label{sec:level1} Result and discussion}

\subsection{\label{sec:level2} Qualitative description of the droplet solidification}

\begin{figure*}
\includegraphics[width=\textwidth]{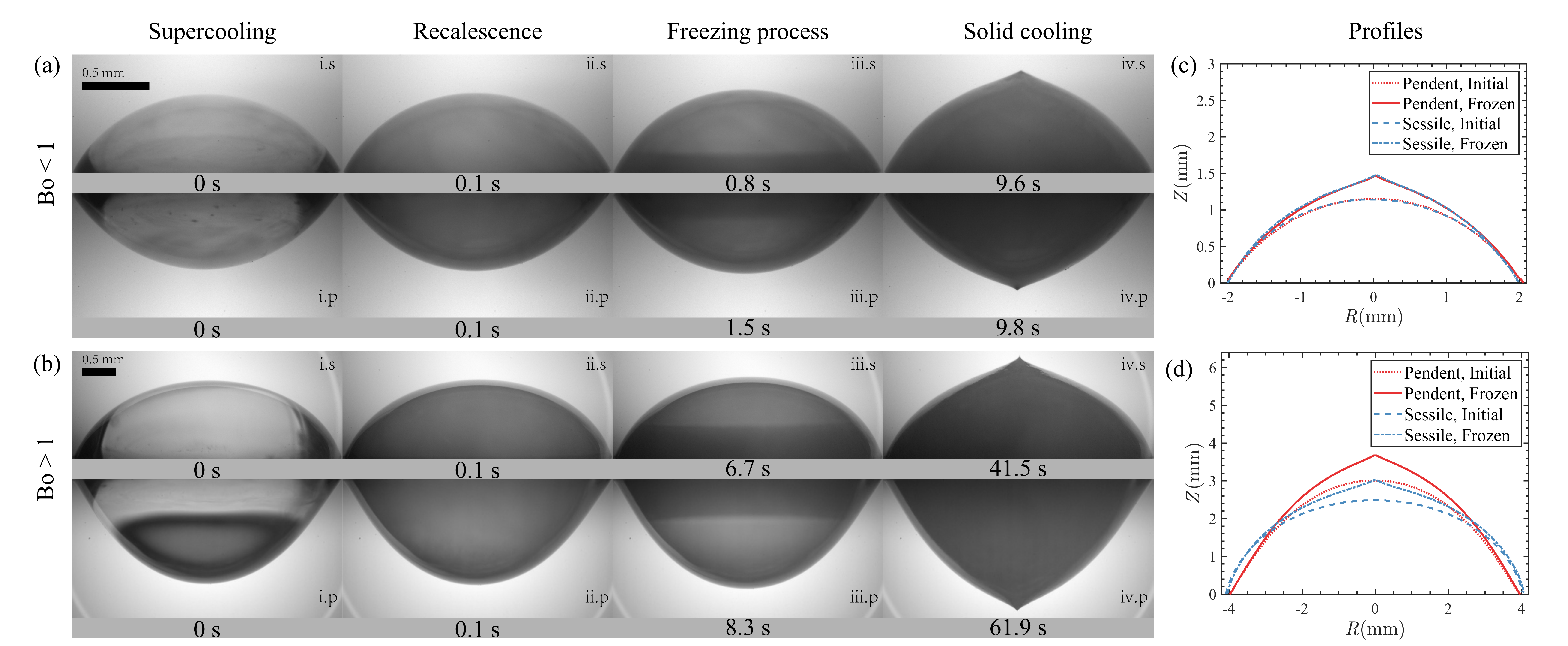}
\caption{\label{fig:seq} Experimental results of pendent and sessile water droplets freezing on a cold substrate; (a-b): side-view snapshots for droplets at different freezing stages; (c-d) initial and final droplet profiles extracted from (a-b) respectively; (a) For $Bo < 1$ case, the droplet base radius is $R_0 = 2$ mm and its initial volume is $V_0 = 8.5 \ \rm\mu L$ ; (b) For $Bo > 1$ case, the droplet base radius is $R_0=4$ mm and its initial volume is $V_0 = 78 \ \rm\mu L$.}
\end{figure*}

As shown in Fig.~\ref{fig:seq}, both sessile and pendent droplets have similar freezing processes, which can be divided into four stages : (1) supercooling, (2) recalescence, (3) freezing, and (4) solid cooling\cite{zhang2018simulation}. During this process an important shape change happens to the droplet, making a sphere-like water droplet turns into an ice cone with a sharp tip. The freezing of water droplets first starts with supercooling. In this process, the mean temperature of the water droplet can decrease down to below the melt temperature. Then the accumulated supercooling degree is released in the recalescence stage, leading to rapid ice crystal formation and usually accomplished within several tens of milliseconds \cite{Jung2012}. After the recalescence stage, the droplet turns into a mixed-phase state consisting of water-ice mixture with temperature equals to the freezing temperature~\cite{Sebilleau2021}. Droplet shape also changes a little bit due to volume expansion. In the freezing stage, the ice layer starts to grow from the wall to the droplet apex, forming a moving ice-water interface within the droplet. During the advancement of this interface, which is also referred to as the "icing front", the volume of the remaining liquid decreases and its shape also changes due to surface tension effects. Finally, a sharp tip is formed at the apex position of the droplet \cite{Marin2014}. In the experiments, we varied the direction of gravity subjected to the droplet by flipping the direction of the cold substrate.  

Figure~\ref{fig:seq} presents the side view snapshots acquired in the experiment of a typical freezing process at different Bond numbers for the same volume droplets at both sessile and pendent states. As shown in Fig.~\ref{fig:seq}(a,b)(i.s-iv.s), when placed on the upper surface of the cold plate, the droplet is usually referred to as a "sessile droplet". For a sessile droplet, the direction of gravity is opposite to the direction of ice growth and tends to flatten the droplets on the contact solid surface. When a droplet is suspended on the lower surface of the cold plate, as shown in Fig.~\ref{fig:seq}(a,b)(i.p-iv.p) the direction of gravity coincides with the direction of ice growth and tends to repel the droplet away from the solid surface, known as “pendent droplet”. The Bond number, which is defined as ${\Delta \rho g D_{eq}^{2}}/{\sigma}$, is a dimensionless parameter used to characterize the level of deformation of droplets in a gravity field by evaluating the importance of gravitational forces compared to surface tension forces, in which $\Delta \rho$ is the density difference {between the liquid and gas phases}, $g$ is the gravity acceleration, $D_{eq}=\left(6 V_0 / \pi \right)^{1/3}$ is the equivalent diameter defined with the droplet initial volume $V_0$ and $\sigma$ is surface tension. When the Bond number is greater than 1, gravitational forces dominate the shape of the free surface, and the droplet is deformed due to gravity. When the Bond number is smaller than 1, the shape of the droplet is controlled by surface tension and its shape is a spherical cap \cite{Schetnikov2015}. 

The initial and final droplet profiles extracted from the experiment snapshots are presented in Fig.~\ref{fig:seq}(c,d). Figure~\ref{fig:seq}(c) compares the results for $Bo < 1$ cases, where the base radius of droplets is $R_0=2$ mm and droplet volume is $V_0 =8.5 \ \rm \mu L$. The results show that the initial shape and the final shape of the sessile droplet and pendent droplet with the same base radius and volume  are very similar in this regime. This comparison indicates that for cases with $Bo < 1$, the shape of the liquid-gas interface for both sessile and pendent droplets remains controlled by surface tension during the icing and the change of gravity direction does not induce a noticeable difference. However, as shown in Fig.~\ref{fig:seq}(d), for cases with $Bo > 1$, both the initial and final shape of droplets shows a noticeable difference. Here the base radius of droplets is $R_0=4$ mm and droplet volume is $V_0 =78 \ \rm\mu L$. Before freezing, the pendent droplet tends to have a larger initial height and a smaller contact angle than the sessile droplet with the same volume. {The droplet thickness before freezing is 2.5 mm and 3.0 mm for the sessile case and the pendent case, respectively, corresponding to a significant difference of 20\%.} This significant difference is a result of different directions of gravity on the free surface of the droplet. The gravity direction changes the pressure distribution inside the droplet: according to Eq.~\eqref{e1}, the pressure outside the droplet being equal to the atmospheric pressure at the droplet cusp, the pressure inside the pendent droplets is lower than in sessile droplets at the droplet cusp due to the hydrostatic pressure distribution inside the droplet. This pressure distribution results in smaller liquid surface curvature at the same vertical distance to the apex. After freezing, the difference in droplet profiles becomes more distinguishable between sessile and pendent droplets. {The droplet thickness after freezing is 3.0 mm and 3.7 mm for the sessile case and the pendent case, respectively, indicating a height difference of 0.7 mm comparing with 0.5 mm for droplet initial shape. The difference in droplet height is further increased and the cone shape is more pronounced for the frozen pendent droplet.}

One further fact that emerged from these observations is that for all cases we have investigated, a similar tip structure has been observed to appear on the apex of the drop at the end of the freezing stage. The gravitational direction obviously changes the initial geometry of large droplets ($Bo > 1$) and results in different frozen geometry as it will be discussed in the following. However, the formation of the icing tip is not affected by gravity direction because at the end of the icing, the  liquid volume  is becoming smaller and smaller with a Bond number becoming rapidly much smaller than 1 so that  the final tip shape is  controlled by  surface tension and independent of any gravity effect.

\subsection{\label{sec:level2} Droplet height}

\begin{figure}[b]
\includegraphics[width=0.90\textwidth]{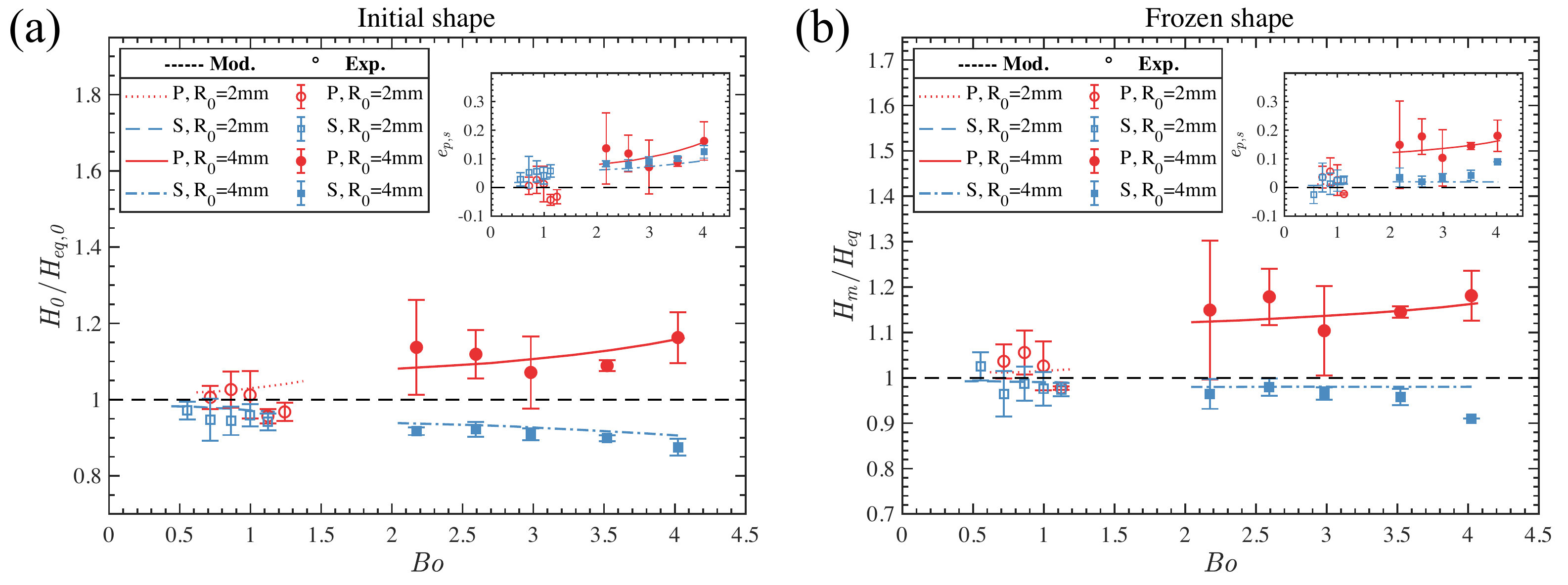} 
\caption{\label{fig:height ratio} Effects of gravity on droplet (a) initial and (b) final shapes for various Bond numbers. The initial $H_0$ and final $H_m$ droplet heights are normalized by their corresponding equivalent height $H_{eq, 0}$ and $H_{eq}$ (see the text), respectively. Vertical bars show $\pm 1$ s.d. The horizontal black dashed line denotes the no-gravity case. The "P" note and red color on the legend mean pendent droplet, while the "S" note and blue color mean sessile droplet. Insert, the deformation strength versus the Bond number.}
\end{figure}

We now turn to the quantitative analysis of the effect of gravity direction on the final droplet shape after freezing. As analyzed in the previous section, the gravity direction has a great influence on the droplet shape. For Bond numbers larger than 1, the height of pendent droplets measured in our experiments [see Fig.~\ref{fig:seq}(d)] are larger than that of sessile droplets for both initial shape and final shape. In addition, the  time it takes to freeze the entire droplet is directly related to the height of the droplet, and the freezing time is also expected to be impacted by the gravity direction. Figure~\ref{fig:height ratio} shows the normalized droplet heights $H_0/H_{eq, 0}$ and $H_m/H_{eq}$ for sessile and pendent droplets as a function of the Bond number. {$H_{eq, 0}$ is the equivalent droplet height defined as the height of a spherical cap with the same volume $V_0$ and based radius $R_0$, while $H_{eq}$ is the droplet height obtained using the model when neglecting gravity.} Both $H_0/H_{eq, 0}=1$ and $H_m/H_{eq}=1$ corresponds to the droplet height when gravity effects are negligible. 
For droplets with  Bond numbers smaller than 1 (droplet base radius $R_0 = 2$ mm), the droplet volume is varied from 4.2 $\rm \mu L$ to 12.2 $\rm \mu L$, and 6.2 $\rm \mu L$ to 18.2 $\rm \mu L$ for sessile and pendent droplets, respectively. For Bond numbers larger than 1 (droplets with base radius fixed to $R_0 = 4$ mm), the droplet volume is varied from 33 $\rm \mu L$ to 83 $\rm \mu L$ for both sessile and pendent droplets. The experiment data points are averaged over a set of repeated experiments, and the corresponding standard deviation is also presented using  error bars in the figure. 

First, the comparison of the results between the experiments and the simple model shows a satisfactory agreement for both the initial (Fig.~\ref{fig:height ratio}(a)) and final  (Fig.~\ref{fig:height ratio}(b)) shapes of the droplet. This indicates that the droplet shape is always controlled by the balance of surface tension and gravity. Indeed, the viscous-capillary time $t_\sigma= \mu D_{eq} /\sigma$ being several orders of magnitude smaller than the freezing time, surface tension  adjusts instantaneously to gravity.

Gravity has an increasing effect on the initial droplet height when increasing the Bond numbers as shown in Fig.~\ref{fig:height ratio}. For Bond numbers smaller than 1, the droplet initial shape and final shape satisfy $H_0\approx H_{eq,0}$ and $H_m\approx H_{eq}$, respectively.
For an initial Bond number larger than one, the gravity direction induces a clear effect on the droplet shape and final height. As long as the Bond number based on the remaining liquid volume to ice is larger than one, a pendent liquid volume is elongated in the gravity direction, while a sessile liquid volume is flattened. 


To quantify the gravity impact on the freezing, we introduce the droplet relative deformation  as
\begin{equation}
e_{p,s}=\frac{|H_{p,s}-H_{eq}|}{H_{eq}}
\end{equation}
where the $p$ and $s$ subscripts represent the pendent and sessile case, respectively. $e_p$ and $e_s$ are reported in the insert of Fig.~\ref{fig:height ratio}(a) and Fig.~\ref{fig:height ratio}(b). Concerning the effect of deformation on the initial shape we observe that the impact of gravity is almost symmetric comparing the sessile and the pendent droplets, and $e_p \approx e_s$. However, when considering the final iced shape, there is a clear difference between the sessile and pendent droplets, and $e_p \gg e_s$. For the sessile droplet icing dilation and gravity deformation are acting in opposite directions and the gravity effect is almost compensated for droplets considered here. For the pendent droplet, the icing dilatation and the shape elongation in the gravity direction are acting in the same way to increase the droplet final height, so that a stronger deformation is observed, as clearly shown in Fig. 3.

\begin{figure}
\includegraphics[width=0.50\textwidth]{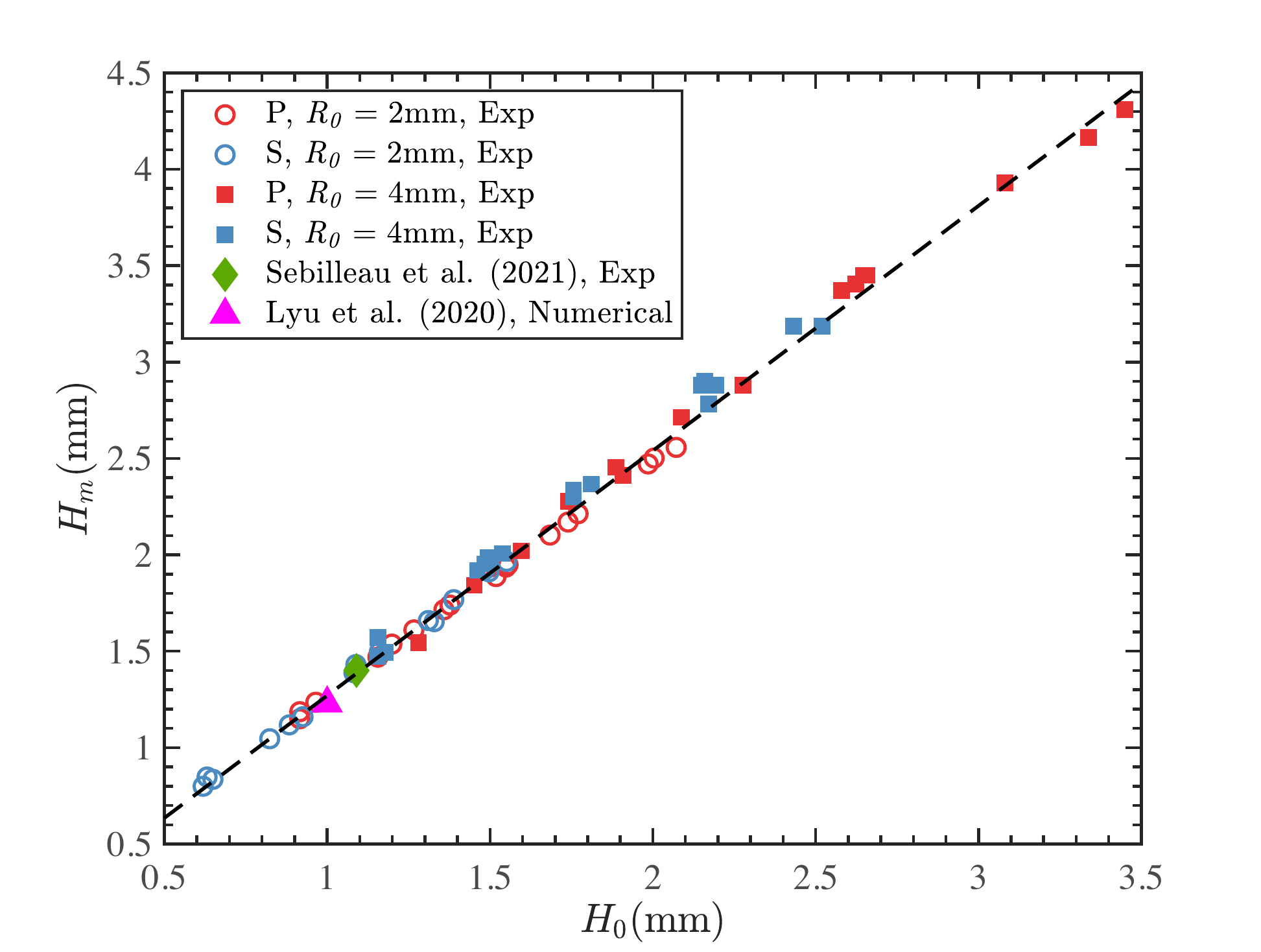} 
\caption{\label{fig:H_m_H_0} Maximum drop height $H_m$ as a function  of $H_0$. The black dashed line '$--$' shows  $H_m = 1.27 H_0$ from relation (\ref{eq:Hm_H0}).}
\end{figure}

We finally compare the final droplet height $H_m$ to the initial droplet height $H_0$ in Fig.~\ref{fig:H_m_H_0}. The experiment results for all the cases: pendent and sessile droplets as well as small and large Bond numbers, are reported. A remarkable linear relation is observed between $H_m$ and $H_0$ as
\begin{equation}\label{eq:Hm_H0}
H_m \approx 1.27 H_0
\end{equation}
 which shows a homothetic translation of the drop height during the icing. As shown in the figure, this relation is in good agreement with both the experiment of Sebilleau et al.~\cite{Sebilleau2021}for a water droplet of initial volume $V_0=8.5 \ \mu$L (Bond number $Bo=0.87$, $H_0=1.09$ mm, $H_m= 1.26 H_0$) and the direct numerical simulation from Lyu et al. \cite{lyu2021hybrid} of a spherical cap water drop with an initial volume $V_0 = 2.09 \ \rm\mu L$ (Bond number $Bo = 0.34$, $H_0=1$ mm, $H_m= 1.23 H_0$).  This vertical translation of the droplet height is dependent on the density ratio between the ice and the liquid so that the pre-factor in Eq. (\ref{eq:Hm_H0}) is varying with the density ratio between the ice and the liquid.
 
\subsection{\label{sec:level2} Freezing front dynamics}

\begin{figure}
\includegraphics[width=0.95\textwidth]{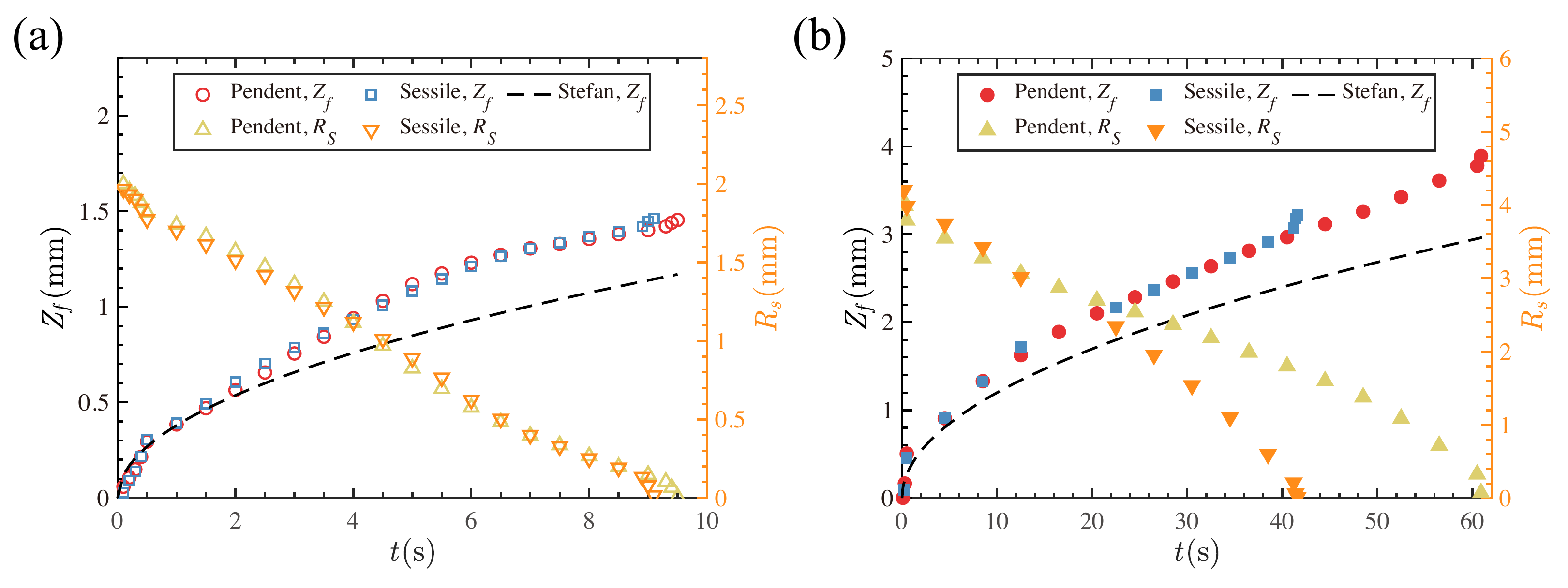} 
\caption{\label{fig:front_evolution} Temporal evolution of the freezing surface contact line $Z_f$  and the freezing front radius $R_s$. (a) small droplets with $Bo = 1.12$; (b) Large droplets with $Bo = 4.02$.  The experimental data are presented with symbols and colors for different gravity directions and front parameters. The solution for the classic Stefan problem $z(t)= 2\lambda \sqrt{\alpha_i t}$, in black dashed line, is determined with the same freezing conditions. }
\end{figure}

 To show how gravity influences the droplet freezing dynamics, Fig.~\ref{fig:front_evolution} reports the vertical position $Z_f$ of the ice-water-air contact line  and the radius of the freezing front $R_s$ (as defined in Fig.~\ref{fig:model}(a)) as a function  of time $t$ for both sessile and pendent droplets at different Bond numbers.   
For Bond numbers smaller than 1, both the sessile droplet and the pendent droplet have the same icing dynamics. As shown in Fig.~\ref{fig:front_evolution}(a) for $Bo = 1.12$ the evolution of $Z_f$ and $R_s$ are very close for the two cases. Due to a liquid spherical cap shape controlled by surface tension, the front propagation and shape evolution for pendent and sessile droplets are similar in the entire freezing process, which further supports the idea of the limited effect of gravity in the freezing of droplets at small Bond numbers. 
\\
However, for Bond numbers larger than 1, the dynamics of the icing are different when comparing sessile and pendent droplets. This is clear in Fig.~\ref{fig:front_evolution}(b) when the case $Bo=4.02$ is reported. Pendent droplets have larger droplet heights than sessile droplets due to gravity-induced deformation, and as a consequence, the total freezing time for a pendent droplet is longer than that of a sessile droplet. Surprisingly, despite a different  freezing time, the position $Z_f$ of the ice-water-air contact line  follows the same evolution for the two cases.  
The time evolution of the freezing front radius $R_s$ indicates that the radii of the freezing fronts also exhibit noticeable difference between sessile and pendent droplets, a direct consequence of a difference in droplet shape between the two cases.
 Fig.~\ref{fig:front_evolution}  also shows comparison with the Stefan solution corresponding to a 1-D front propagation as
 \begin{equation} \label{eq:Stefan}
z(t)= 2\lambda \left(\alpha_i t\right)^{0.5}
\end{equation}
where $\alpha_i$ is the thermal diffusivity of the solid phase and  $\lambda$ is a parameter determined by the Stefan conditions \cite{lyu2021hybrid}.
 {The Stefan solution can describe the experimental evolution at the beginning of the icing for both small and large Bond numbers. But, a clear difference is then observed during the icing. As observed, the freezing is faster than the Stefan 1-D solution for both small and large $Bo$ cases. The departure happens at about 2 seconds after the initiation and gradually deviates in the later stage of the freezing process.} As observed, the freezing is faster than the Stefan 1-D solution.

\begin{figure}
\includegraphics[width=0.50\textwidth]{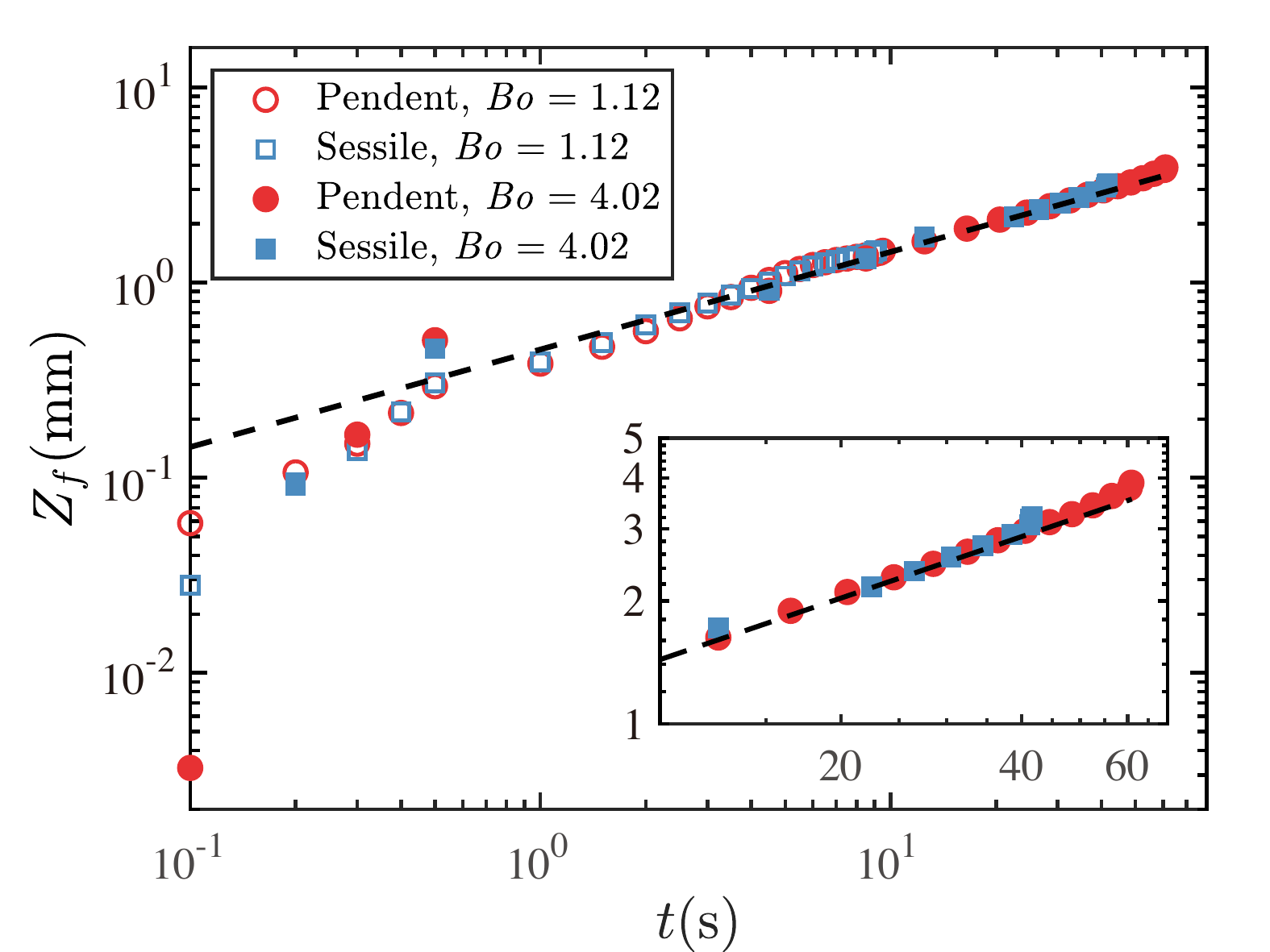} 
\caption{\label{fig:front_evolution_nor} Front height $Z_f$ as a function of freezing time $t$, corresponding to the case previously presented in Fig.~\ref{fig:front_evolution}.  The red and circular marker represents the pendent droplets, while the blue and square marker represents the sessile droplets. The dashed lines with slopes of 0.5 is to guide the eye. Insert shows a zoom of the front propagation at the end of freezing processes for droplets whose Bond number is 4.02.}
\end{figure}
For a detailed inspection of the difference with the Stefan evolution, the evolution processes of $Z_f$ are reported using a log-log plot in Figure~\ref{fig:front_evolution_nor}.
As shown, $Z_f$ follows the same power law evolution as given by the 1-D Stefan solution (\ref{eq:Stefan}) for both sessile droplets and pendent droplets:
\begin{equation}
Z_f \propto t^{0.5}
\end{equation}
This result is consistent with most of the results from the literature. However, we observe here that the pre-factor is larger than the one given by the Stefan solution  (\ref{eq:Stefan}). This difference is clearly identified from the plot. Very interestingly, all the cases considered, small and large Bond numbers as well as sessile and pendent droplets, are following the same evolution. Thus,is faster icing dynamics cannot be attributed to any  gravity effect on the droplet shape. {Such difference was recently discussed in Refs.~\cite{thievenaz2019solidification,lyu2021hybrid,Sebilleau2021}. This fast evolution can be partially attributed to the constriction of the heat flux in the droplet during the freezing process. The based radius in contact with the cold surface remains constant while the icing front surface is decreasing. In addition, the ambient humidity has been shown to significantly impact the icing kinetics because of either condensation at the droplet surface reducing the ice front propagation, or evaporation at the droplet surface increasing the solidification process. Here the use of nitrogen to prevent frost formation on the freezing substrate results in some evaporation at the droplet surface. Thus the faster icing kinetics observed here is attributed to a combination of droplet geometry and water evaporation at the droplet surface.} 

 \subsection{\label{sec:level2} Droplet freezing time}

\begin{figure}\includegraphics[width=0.90\textwidth]{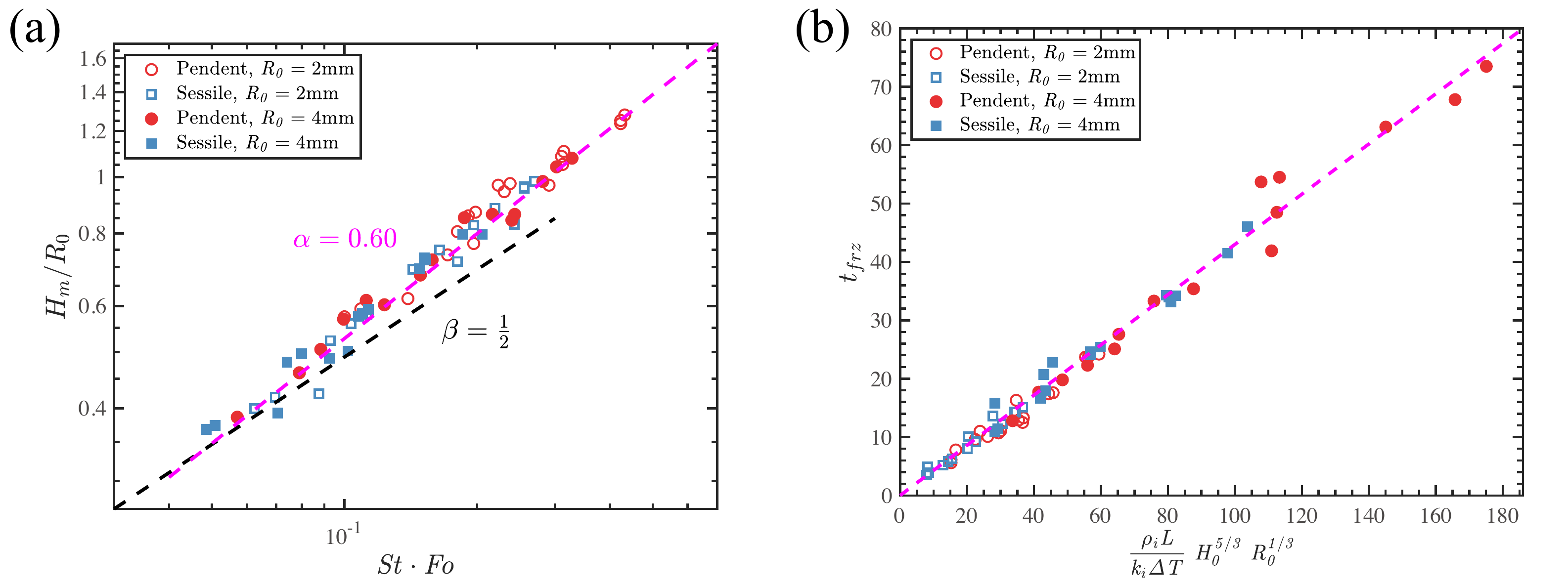} 
\caption{\label{fig:scaling} Relation between the droplet freezing time $t_{frz}$ and droplet shape. (a). The final droplet height $H_m$ is normalized using $R_0$ to eliminate the influence of different base radii while $t_{frz}$ is normalized using the Stefan and Fourier numbers according to Eq. 2. The dashed line with slopes of 0.60 corresponds to Eq. 7 with $C = 2.1$ and the dashed line with slope 0.5 is to guide the eye. {(b). Measured freezing time versus predicted freezing time. The dash line with a slope of 0.43 represents the value predicted with equation 8.}}
\end{figure}

 We finally consider the freezing time $t_{frz}$ i.e.  the required time  to completely froze the droplet. From the 1-D Stefan solidification solution given by Eq. (\ref{eq:Stefan}), $t_{frz}$ can be related to the final drop height  as
\begin{equation}\label{eq:Hm0}
H_m = 2 \lambda \left(\alpha_i t_{frz} \right)^{0.5}
\end{equation}
When the temperature of the liquid equals the freezing temperature as a result of the recalescence and when the latent heat $L$ is much larger than the sensible heat contribution, it yields $\lambda=\sqrt{{c_i \Delta T}/{2L}}$ where $c_i$ is the ice heat capacity. Introducing the Stefan number $St={c_i \Delta T}/{L}$ and the Fourier number $Fo={k_i t_{frz}}/{c_i \rho_i R_0^2}$, with $\rho_i$ the ice density, to normalize the freezing time $t_{frz}$ in Eq. (\ref{eq:Hm0}), the normalized final drop height is expected to evolve with the normalized freezing time as
\begin{equation}\label{eq:Hm}
H_m/R_0 \propto St^{1/2} \ Fo^{1/2}
\end{equation}
Figure~\ref{fig:scaling}(a) reports the evolution of $H_m/R_0 $ as a function of $St \cdot Fo$ for all the experiments. The droplet freezing times and initial heights are obtained in 62 independent freezing experiments with ten different Bond numbers, for sessile and pendent droplet. The experimental points all collapse on a single curve showing no significant effect of gravity. A clear power law is also observed but with an exponent $\alpha = 0.60$ larger than the one given by the Stefan solution according to Eq. (\ref{eq:Hm}).
From Fig. \ref{fig:scaling}(a), a relation can be proposed to relate the height of the iced drop $H_m$ and the freezing time as
\begin{equation}\label{eq:Hm_1}
H_m/R_0 = C \ St^{0.6} \ Fo^{0.6}
\end{equation}
with the value $C\approx 2.1$ obtained by fitting the experimental points. 
The difference with the 1-D Stefan problem is related to the geometrical constriction of the liquid remaining to ice when pushed by the dilatation induced by the water solidification, but also to the operating conditions related to the recalescence stage and water evaporation at the droplet surface in the ambient Nitrogen\cite{Sebilleau2021}. Combining relation (\ref{eq:Hm_1}) with relation (\ref{eq:Hm_H0}) between $H_m$ and $H_0$, the freezing time can be expressed as a function of the initial drop height and base radius as
\begin{equation}\label{eq:t_frz}
t_{frz} \approx 0.43 \ \frac{\rho_i L}{k_i \Delta T} \  H_0^{5/3} \ R_0^{1/3}
\end{equation}
As shown in Fig.~\ref{fig:scaling}(b), this simple relation allows determining the freezing time as a function of the prior known parameters. In particular the freezing time is shown to increase as $R_0^{1/3} \ H_0^{5/3}$ for both sessile and pendent droplets, large and small Bond numbers. Thus, the longer time required to freeze a pendent droplet compared to a sessile droplet of the same volume is completely taken into account by considering their initial geometry.


\section{Conclusions}

The effect of gravity on water droplet freezing on a cold solid surface has been investigated. The experiments with pendent and sessile droplets have been conducted for Bond numbers smaller and larger than unity. For Bond numbers smaller than unity, the freezing process of sessile and pendant droplets is not affected by gravity because the shape of the liquid part of the droplet is controlled by surface tension during the entire freezing process. For Bond numbers larger than unity, gravity modifies the initial shape of the droplet: sessile droplets are flattened while pendents droplet are elongated. It results in a longer time to ice for a pendent droplet than for a sessile droplet. Despite these differences in the initial droplet shape, several remarkable similarities have been found for all the configurations, pendent and sessile droplets, small and large Bond numbers.
(i) The final height of a frozen droplet is found to be linearly proportional to its initial height $H_m=1.27 H_0$.
(ii) The time evolution of the ice-liquid-air contact line is found to follow the classical power-law $t^{0.5}$, thus noticeably faster than the Stefan 1-D icing front propagation because of the effect of the initial recalescence and water evaporation at the droplet surface.
(iii) As a consequence, the time to freeze a droplet is faster than predicted by the Stefan model and it has been found to evolve as $t_{frz} \propto H_0^{5/3} \ R_0^{1/3}$. Thus, the initial geometry of the droplet is completely controlling the icing process.


\begin{acknowledgments}
This work was financially supported by the Natural Science Foundation of China under Grant Nos. 11988102 and 91852202, the National Key R\&D Program of China under Grant No. 2021YFA0716201 and Tencent Foundation through the XPLORER PRIZE. We are grateful to Yuki Wakata for assistance with the code, and Mingbo Li, Ziqi Wang, and Lei Yi for helpful conversations. 
\end{acknowledgments}

\appendix*

\section{\label{app:A}Description on the simple model}

{Based on previous studies, we have proposed a simplified model here. The geometry calculation in the model is first inspired by Ref.~\cite{sanz1987influence,Anderson1996}, and the heat transfer calculation is optimized by correcting the heat flux in the later stage of freezing. The detailed descriptions on this model are given in the following parts.}

\subsection{\label{sec:level2} Liquid shape}
For a droplet in the gravity field, the droplet shape is determined by the balance of the gravity and surface tension. The gravity causes a downward positive pressure gradient inside the droplet. However, with different deposition directions, the gravity has different effects on the droplet shape. For any point on a free surface of the water droplet, the curvature $\kappa$ can be described by the Young-Laplace equation as shown below:
\begin{equation}
\label{e1}
\kappa=\kappa_{0} \pm \frac{\rho_f g}{\sigma}z.
\end{equation} 
Here, the choose of sign $\pm$ is related to the direction of gravity, $'+'$ sign for sessile droplets and $'-'$ sign for pendent droplet. $\rho_f$ is the liquid density, $\kappa_{0}$ refers to the curvature at the droplet apex, $z$ is the vertical distance to the droplet apex and $\sigma$ is the surface tension for the water-air interface. Normalized by the capillary length $l=\sqrt{\frac{\sigma}{\rho_f g}}$, then Eq.~\eqref{e1} can be rewritten as:
\begin{equation}
\label{laplace-y-nor}      
\kappa^*=\kappa_{0}^* \pm z^*.
\end{equation}
with $\kappa^*=\kappa l$, $z^*=z / l$. As shown in Fig.~\ref{fig:model} (a), integrating Eq.~\eqref{laplace-y-nor} with the arc length $s$ along with the liquid-gas interface from the droplet apex to the droplet base, the equation can be solved by introducing the tangential angle $\phi_f$ between arc length and horizontal line\cite{Cai2020}.

\subsection{\label{sec:level3} Freezing stage}
According to the conservation of mass at the solidification front $Z=Z_f$. The radius and remaining volume follows the relation as:
\begin{align}
    &\frac{dR_s}{dZ}=\textrm{tan}(\frac{\pi}{2}-\phi_f+\phi_i),\label{2}\\
    &\frac{dV}{dZ}=-\pi \rho R_s^{2} ,\label{3}    
\end{align}
The sketch of the drop is shown in Fig.~\ref{fig:model} (a), where $R_s$ is the radius of the front, Z is the axial coordinate, V is the remaining liquid drop volume, $\phi_f $ is the angle between the flat solidification front and the liquid interface at the trijunction point, $\phi_i$ is the dynamic growth angle referring to the angle between the tangential lines of gas-liquid interface and gas-solid interface and $\rho=\rho_s / \rho_l$ is solid to liquid density ratio. The initial conditions are:
\begin{equation}
    R_s(0)=R_0, V(0)=V_0, \label{4}
\end{equation}
where $R_0$ is the radius of droplet base, and $V_0$ is the initial volume of the droplet. When the above equation are normalized by $r_s=R_s/R_0, z=Z/R_0$ and $v=V/R_0^{3}$, we can obtain the dimensionless differential equations at the icing front $z=z_f$ as:
\begin{align}
    &\frac{dr_s}{dz}=\textrm{tan}(\frac{\pi}{2}-\phi_f+\phi_i), \label{e5}\\
    &\frac{dv}{dz}=-\pi \rho r_s^{2}, \label{e6}    
\end{align}
with the appropriate initial conditions:
\begin{equation}
    r_s(0)=1, v(0)=v_0. \label{7}
\end{equation}
The $\phi_f$ is determined by the shape of the remaining liquid droplet, which can be derived from Eq.\eqref{e1} with given base radius $r=r_s$ and remaining volume $v$. {As shown in previous studies, the contact angle between frozen and liquid water can be influenced by the conditions at the contact line\citep{thievenaz2020freezing}. With the flat ice front assumption, the ice profile growth not always follows the tangential direction of the liquid-gas interface\citep{Anderson1996}.} When the liquid contact angle $\phi_f $ exceeds the critical receding angle, the dynamic growth angle $\phi_i$ is nonzero, which is determined by the conditions at the trijunction line. The contact line slips at a speed $U_s$, which follows the geometrical relation as shown in the figure:
\begin{equation}
    U_s=U_f(\frac{1}{\textrm{tan}(\phi_f-\phi_i)}-\frac{1}{\textrm{tan}(\phi_f)}), \label{eq_Vs}
\end{equation}
where $U_f$ is the solidification front advancing speed. The slip speed also associated with the contact angle, and the relation between them are given by\cite{Lu2022,Anderson1996}:
\begin{equation}
\label{vs_relation}
U_s=\left\{
\begin{aligned}
&\frac{\eta(\phi_r-\phi_f)}{\phi_f}&(\phi_f \ge \phi_r), \\
&0&(\phi_f < \phi_r),
\end{aligned}\right .
\end{equation}
{where $\eta$ is the characteristic slip velocity related to the solidification rate at the freezing front and $\phi_r$ is the critical receding angle.}
The energy balance at the freezing front can be described as:
\begin{equation}
    \pi r_s^{2} \rho_s \frac{dz_f}{dt} = \int \frac{k_i \Delta T}{z} dA, \label{energy}
\end{equation}
{Though the flat solidification front is observed in some experiments at the beginning of the icing, the ice front is not flat near the interface and at the end of the icing the front is more like a spherical cap\cite{Marin2014}. Considering the actual shape of the solidification front, in the later freezing process, the actual area that brings heat away from the remaining liquid is increased when compared with a flat interface. For simplicity, two stages approximation is usually applied to models to simulate the unique change in front shape\cite{Nauenberg_2016}. In this model, the heat transfer area is regarded as a spherical cap once the ice layer thickness exceeds $h$, the height of the corresponding spherical cap. The critical height is given by $h={r_s}(1/{\rm{sin} \phi_c}-\rm{cos} \phi_c)$, where $r_s$ is the ice front radius and $\phi_c$ is the base angle of the spherical cap $\phi_c = \pi/2 -\phi_f$}. Thus the effective heat transfer area can be expressed as:
\begin{equation}
    \label{area}
    A=\left \{
    \begin{aligned}
    &\pi r_s^{2}&(z_f < h),\\
    &2\pi r_c^{2} (1-\textrm{sin}(\phi_f))&(z_f \geq h),
    \end{aligned}\right .
\end{equation}
where $r_c=r_s/ \text{cos} (\phi_f) $ is the radius of curvature for the spherical cap. Thus the relation between the solidification front height and the freezing time can be given by:
\begin{equation}
    \label{dzdt}
    \frac{dz_f}{dt}=\left \{
    \begin{aligned}
    &2\frac{k_i \Delta T}{z_f\rho_i L_m}& (z_f < h),\\
    &\frac{2k_i\Delta T}{r_s \textrm{cos}(\phi_f)\rho_i L_m}\text{ln} \left(\frac{z_f}{z_f+r_c (\text{sin}(\phi_f)-1)} \right )&(z_f \geq h).
    \end{aligned}\right .
\end{equation}

Combining these equations, we can obtain a simple model that can be used to describe the dynamic process of droplet freezing in different gravity conditions. With a given set of conditions droplet  volume, base radius, and direction of gravity, the initial droplet shape can be calculated. The equations give different solutions for the initial droplet profile with different gravity directions. Then the initial droplet profile, as well as the droplet volume and base radius, are used as input parameters of the calculation for freezing. In the freezing stage, the droplet can be divided into two parts: the ice part and the remaining liquid. Both of them are coupled with the evolution of the liquid volume $v$. The shape of the ice part is determined by the previously calculated evolution of the height of the freezing front and the radial distance $r_s$ of the ice-water-air contact line, while the shape of the remaining liquid is updated at each time step of the calculation according to the remaining liquid volume $v$ and front radius $r_s$. The prediction of the critical receding angle $\phi_r$ is based on the inflection point in the droplet profile\cite{Anderson1996}. Thus we can regard $\phi_r$ as a variable related to the initial contact angle $\phi_0$ as $\phi_r=\xi \phi_0$ for different droplets in the calculation. With the conditions confirmed by the initial shape and prediction of the inflexion point, it is find that $\rm\eta=1.1 \rm mm/s$ is a reasonable value for all the experiment cases, and $\xi=0.5$ agree well for most cases while $\xi=0.8$ for the pendent droplet with a base radius $R_0 = 4 \rm mm$ contact with the substrate. The Eq.~\eqref{e5},\eqref{e6},\eqref{dzdt} are integrated using a fourth-order Runge-Kutta scheme. For the stop criteria of the numerical resolution, the integration procedure stops at $v=\epsilon$ ($\epsilon=1\times10^{-6}\ll 1$ in this study).


\providecommand{\noopsort}[1]{}\providecommand{\singleletter}[1]{#1}%

\end{document}